# Error Source Sensitivity Analysis in Model-Based Coherence Scanning Interferometry for Thin Film Metrology

**LIXUAN XU,**[1,2] **CHENG CHEN,**[2,*] **AND RONG SU**[2,3]

[1] *School of Microelectronics, Shanghai University, Shanghai 201800, China*
[2] *Shanghai Institute of Optics and Fine Mechanics, Chinese Academy of Sciences, Shanghai, 201800, China*
[3] *surong@siom.ac.cn*
[*] *chengchen2020@siom.ac.cn*

**Abstract:** In semiconductor manufacturing processes, silicon dioxide films are commonly used as barrier layers, insulating layers, and protective layers. Coherence scanning interferometry (CSI) offers thin film thickness measurements with a millimeter-scale field of view and micrometer-scale lateral resolution. When the film thickness is less than the coherence length of a CSI system, a model-based film thickness measurement method is typically employed, which relies on a priori information about the thin film and the instrument. This study quantitatively analyzes how the accuracy of a priori information would affect the accuracy of thickness measurement when using a model-based CSI method. The influence factors include camera noise, numerical aperture (NA), pupil apodization, light source spectrum, and thin film refractive index. A series of $SiO_2$/Si thin films with varying thicknesses are analyzed by combining simulation and experimental approaches. The results reveal that the accuracy of thickness measurements exhibits varying sensitivity to different a priori information. The refractive index of the thin film is identified as the most sensitive source of error, where 1% deviation in refractive index may cause 1% relative thickness error, whereas 5% deviation in NA results in less than 1% relative thickness error. The simulation and experimental results show good agreement, validating the correctness and effectiveness of this study.

## 1. Introduction

Transparent films play a critical role in silicon-based integrated circuits (ICs) and other devices [1,2]. For instance, in micro-electromechanical systems (MEMS), silicon dioxide films are commonly employed as structural or sacrificial layers [3]. In the manufacturing process of film structures, precise characterization of film uniformity and thickness, is important to optimize process parameters and minimize defects. Film measurement and inspection can occur in-process or at the final stage, requiring high accuracy and speed. For example, in a chemical-mechanical polishing process, monitoring film thickness is essential to precisely determine the polishing termination point. [4].

A range of techniques, including mechanical, electrical, and optical methods, can be used to measure thin film thickness. Atomic force microscopy (AFM) [5] and stylus profilometry [6] provides high-resolution measurements. However, their application is limited by the need for a distinct step between the film and substrate. This limitation is particularly challenging for dielectric films on similar substrates with masking during deposition or post-deposition etching. Furthermore, stylus methods can damage delicate film surfaces. While electron microscopy, including SEM [7] and TEM [8], offers unparalleled resolution for surface and internal layer imaging, its application is limited by the need for extensive sample preparation and a vacuum environment.

Optical techniques, including spectral reflectometry [9], spectral-domain interferometry [10], spectroscopic ellipsometry [11], and coherence scanning interferometry (CSI) [12], offer non-destructive, high-throughput, and often in-situ characterization of thin film structures. SR measures thin film thicknesses and refractive indices by analyzing the self-interference spectrum from multiple reflections within the film under broadband light illumination, while spectroscopic ellipsometry measures the thin film-induced polarization change under broadband illumination. However, both techniques suffer from low lateral resolution due to the large illumination spot, making the combination of high lateral resolution and high throughput time-consuming or complex to achieve [13].Spectral-domain interferometry operates similarly to spectral reflectometry but incorporates a reference mirror, adding phase information to the spectral signal. Despite this enhancement, spectral-domain interferometry shares the same limitations as spectral reflectometry [14]. In contrast, CSI excels by achieve sub-micrometer lateral resolution with nanometer accuracy across the entire field of view, while also providing sub-nanometer resolution for surface topography profiling [15].

As for the film measurement with CSI, the algorithmic approach behind varies depending on the film thickness. When the film thickness is thick [Fig.1(b)], typically exceeding the coherence length of CSI, two distinct peaks corresponding to reflections from the top surface and the substrate can be identified. Therefore, the positions of these peaks can be localized, enabling the determination of both the film thickness and the surface topography [16]. As the film thickness approaches the submicrometer level [Fig.1(c)], signal peaks from the top surface and the substrate overlap, and so they cannot be separated well. The peaks from the top surface and the substrate converge, making them difficult to distinguish clearly. Most research on film measurement algorithms focuses on this scenario.

Kim et al [17] have introduced a phase model for frequency-domain analysis of CSI signals and utilized a nonlinear least-squares method to simultaneously estimate surface topography and film thickness parameters. Colonna de Lega et al [18] and Fay et al [19] have employed signal library-based or model-based analysis methods to determine film thickness and surface topography. In these approaches, each pixel-wise CSI signal is compared to entries in the library, with the best matching signature providing the corresponding thickness values. Mansfield [20] has developed the helical conjugate field (HCF) model, which represents a topographically defined helix modulated by the electrical field reflectance, specifically for measuring thin film thickness. Along with his colleagues [21], they have simplified the computation by using the first-order Taylor expansion of the HCF. Other studies have used the Fourier magnitude of pixel-wise CSI signals as the measurement signal in SR technique to determine film thickness [22]. Additionally, some researchers have directly compared the measured signal with an estimated signal, simultaneously treating film thickness and surface topography height as unknown parameters [23].

All the methods mentioned above, whether based on the nonlinear least-squares approach or model-based techniques, fundamentally rely on either a phase model or a signal model. Consequently, developing an accurate CSI signal model that closely matches the experimental signals for the film structure is crucial.

Generally, the pixel-wise CSI signal model starts with numerical aperture (NA) equals zero [24, 25]. Subsequently, models that accounts for both spatial and temporal coherence is proposed [26-30]. De Groot et al [31] have introduced a pixel-wise signal model that considers factors such as NA, apodization, and the spectral distribution of the light source. Based on this model [31], Dong et al [32] have investigated how the fluctuation ranges of phase and amplitude curves, when using model-based methods, are influenced by parameters of the model. Lin et al. [33] and Kiselev et al. [34] have conducted similar sensitivity analyses.

However, these sensitivity analyses are limited to the pixel-wise CSI signal signatures and do not cover numerical and experimental investigations of how errors in the instrumental parameters of CSI and the material properties of the film structure would affect the library matching results. Deviations between the actual and nominal values of instrumental and material parameters can introduce undesirable errors. For example, although interferometric microscope objectives are designed with a nominal NA, the actual NA may differ and is often undisclosed. Nevertheless, CSI signal libraries are typically generated based on the nominal NA rather on the actual value. A quantitative assessment of the influence of these deviations is essential to identify the critical parameters requiring calibration, thereby improving measurement accuracy for specific tasks.

In this paper, we quantitatively investigate the impact of errors in CSI instrumental and material parameters on the model-based CSI method, including NA, pupil apodization, the spectral distribution of the light source, camera noise levels, and film refractive index. Numerical analyses are validated through experimental results.

The paper is organized as follows: Section 2 revisits the details of the CSI signal model, thin film model, and the algorithmic flow of the model-based analysis method. Section 3 presents simulations of thickness measurement deviations and relative deviations across various parameters for different single-layer thin film structures. Section 4 discusses the experimental results and compares them with the simulation outcomes. Section 5 summarizes the results.

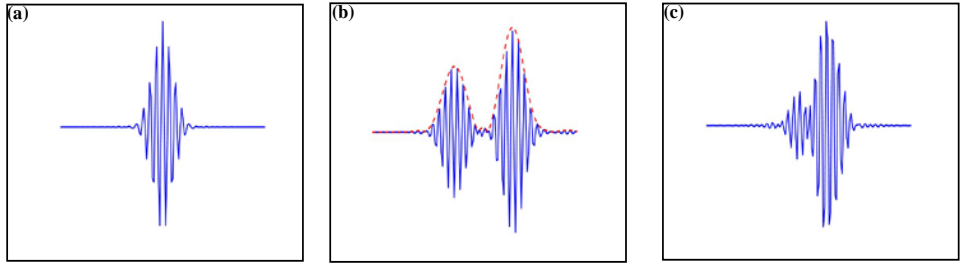

Fig. 1 CSI signals for different sample. (a) signal for bare surface (no film) (b) signal for thick film (over 1μm) (c) signal for thin film.

## 2. Principle of model-based CSI film metrology

A typical CSI setup includes a broadband light source, Köhler illumination optics, an interference objective, and a mechanical scanner. With a spatially extended, spectrally broadband illumination, interference fringes only occur in a small region around the surface along the axial direction, depending on the coherence length of the source and NA of the microscope objective. The images that contain interferograms are sequentially recorded by a camera from a smooth continuous scan of the objective in the z-direction [12].

In the model-based method for thin film measurement, it is essential to model the CSI signal with high accuracy. This process includes developing accurate models for both the CSI instrument and the thin film structure.

### 2.1 Theory

The pixel-wise signal is an integral over all points in the microscope objective pupil plane and over all wavelengths for the ray bundle contributions [31],

$$I(z) = \int_0^\infty \int_{\psi_1}^{\psi_2} g(k,\psi,z) P(\psi) s(k) \sin\psi \cos\psi \, d\psi dk \qquad (1)$$

where $z$ denotes the scanning positions, $k = \frac{1}{\lambda}$ is the wavenumber at an illumination wavelength $\lambda$, $\psi$ is the incident angle within the range of $[\psi_1, \psi_2]$, $s(k)$ denotes the optical spectrum distribution of the light source, $P(\psi)$ is the intensity distribution in the pupil plane of the objective, with the implicit assumption of rotationally symmetric pupil apodization, $g(k,\psi,z)$ denotes the interference signal for a single ray bundle at incident angle $\psi$,

$$g(k,\psi,z) = |r_r(k,\psi)|^2 + |r_s(k,\psi)|^2$$
$$+2\Re\begin{Bmatrix} r_r(k,\psi)^* r_s(k,\psi) \\ exp[j4\pi k(h_s - z)\cos\psi + j\Delta(k,\psi)] \end{Bmatrix} \quad (2)$$

where $\Re$ is the operation for taking the real part, $r_r$ denotes the complex reference reflection coefficient, including both the beam splitter and the reference mirror, $r_s$ denotes the complex reflection coefficient of the thin film structure, including, e.g., the transmission of the beam splitter, $h_s$ represents the height of the upper surface of the thin film structure, * is the complex conjugate operation, and $\Delta$ is an additional phase term associated with the chromatic dispersion of the CSI instrument.

The materials of the thin film structure may exhibit varying properties with respect to polarization. In this study, we limit our analysis to homogeneous and isotropic thin film structures, as more complex film structures can be analyzed using a similar approach. The complex reflection coefficient for the multi-layer thin film structure is given as [35]

$$r_s(k,\psi) = \frac{\gamma_0(k,\psi) - Y(k,\psi)}{\gamma_0(k,\psi) + Y(k,\psi)} \quad (3)$$

where $\gamma_0$ is the medium of admittance of the ambient environment, which in this case is air, and $Y$ denotes the surface admittance of the film structure.

## 2.2 Model-based film metrology

In this work, our analysis—comprising both simulations and experiments—focuses on the $SiO_2/Si$ thin film structure. Examples of CSI signals for the $SiO_2/Si$ thin film structure at varying film thicknesses are shown in Fig. 1.

Figure 2 illustrates the algorithmic flow of the model-based analysis method, which is divided into three main stages: signal modeling, library generation, and signal processing. Signal modeling incorporates the instrumental parameters, such as NA, spectral distribution of the light source, and the film parameters, such as layer structure, top-layer surface topography, refractive index of the film and substrate, and incident angles within NA. Library generation involves creating a series of CSI signals at various layer thicknesses. In the signal processing, the measured CSI signals are compared with the generated signal library to determine the best match for the film thickness.

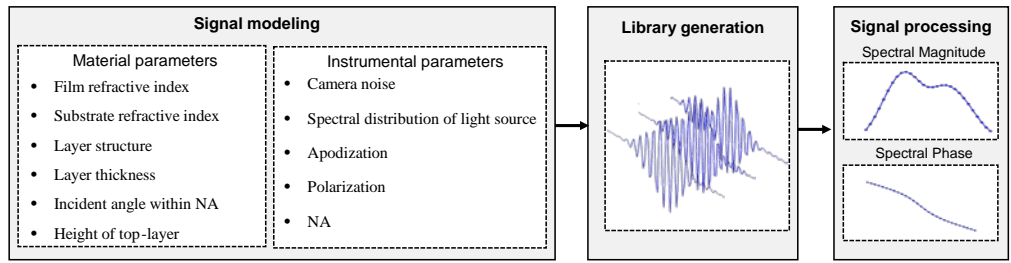

Fig. 2 Workflow of a model-based analysis method for measuring transparent thin film thickness.

A key aspect of the process is ensuring consistency between the parameters used in the signal library and the actual parameters of the measured CSI systems. Deviations between these parameters can lead to inaccuracies in film thickness measurements.

However, preparing a calibrated standard film sample for each specific measurement task is expensive and burdensome, making it challenging to quantify the impact of such deviations on accuracy. To address this, sensitivity analysis is crucial for assessing the influence of these deviations on film measurement accuracy. This approach can help identify the critical parameters that need calibration, ultimately improving the overall measurement accuracy.

## 3. Sensitivity analysis by numerical simulation

First, a series of signal libraries for different film thicknesses are generated using the nominal instrumental and film parameters. Next, a new signal is generated with deviated instrument and film parameters to simulate the measured signal. Finally, the measured film thickness is extracted using a model-based method. During $N$ repeated measurements in the presence of random noise (camera noise, random vibration, etc.), the statistical mean value of the thicknesses measured of repeated measurements is

$$\bar{T}_m = \frac{1}{N}\sum_{i=1}^{N} T_m^i \tag{4}$$

where $T_m^i$ denotes the film thickness measured using the model-based method of the ith measurement. The thickness deviation $\delta_T$ is defined as

$$\delta_T = \bar{T}_m - T_n \tag{5}$$

where $T_n$ denotes the nominal thickness. And the standard deviation of the thickness deviations $\sigma_T$ is defined as the statistical standard deviation of thickness deviations of repeated measurements.

$$\sigma_T = \sqrt{\frac{1}{N-1}\sum_{i=1}^{N}(T_m^i - \bar{T}_m)^2} \tag{6}$$

And the relative thickness deviation $\varepsilon_T$ is defined as the ratio between $\delta_T$ and $T_n$, and the relative standard deviation $\varepsilon_\sigma$ is defined as the ratio between $\sigma_T$ and $T_n$.

All simulations are performed using three commonly used NA values (0.3, 0.4, and 0.55), and five film thicknesses (100, 300, 500, 700, and 1000, unit in nm). A series of CSI signals are generated using a Gaussian-distributed light source spectrum centered at approximately 0.575 μm, with a full width at half maximum (FWHM) of 0.12 μm. The scanning interval is set to one-eighth of the center wavelength. It is important to note that the analysis follows the one-at-a-time (OAT) sensitivity analysis approach, where each parameter is varied individually while all other parameters remain nominal. This method isolates the independent impact of each variable by leveraging the variable control technique.

*3.1 Camera noise*

In this study, we assume the camera noise follows a Gaussian distribution. Figure 3 shows the CSI signals and their Fourier magnitudes for both a noise-free condition and a signal-to-noise ratio (SNR) of 40 dB, with a NA of 0.55 and a film thickness of 500 nm. It is evident that even low levels of camera noise could introduce significant errors in the Fourier magnitude.

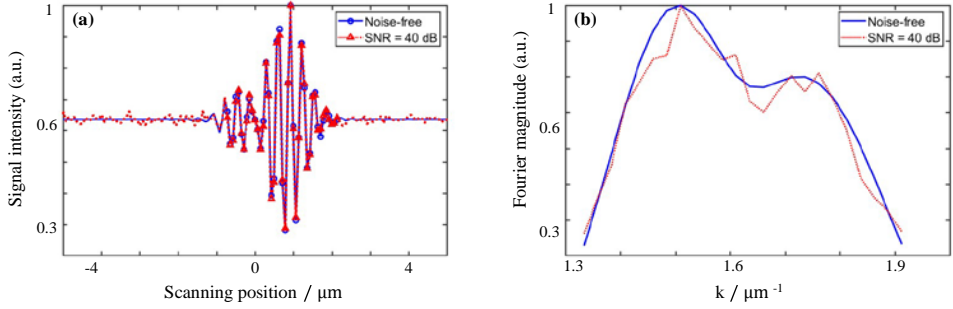

Fig. 3 CSI signals of a 500 nm SiO$_2$/Si film structure at NA=0.55 (a) and their Fourier magnitudes (b) with and without camera noise. Noise is simulated for SNR=40 dB.

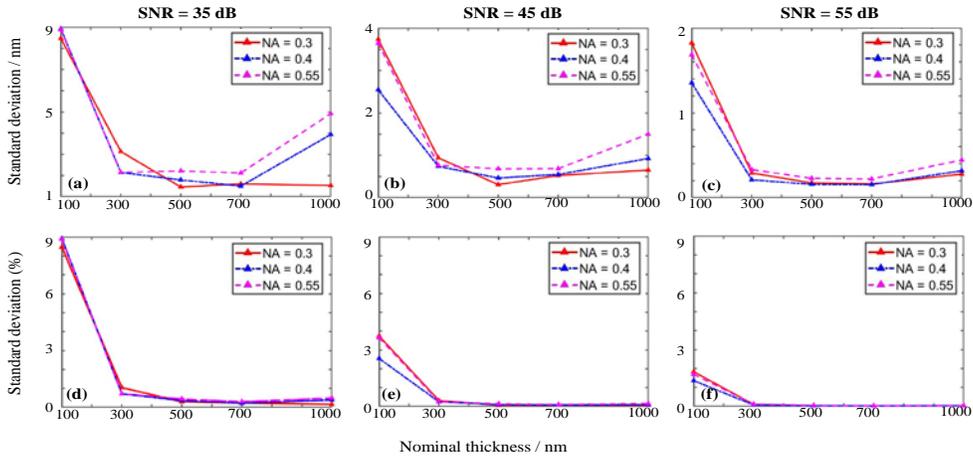

Fig. 4 Standard deviation and relative standard deviation of the absolute thickness deviations across different NAs, SNRs, and film thicknesses

Figures 4(a), (b), and (c) illustrate the variation of $\sigma_T$ with respect to the nominal thickness $T_n$ at SNR levels of 35dB, 45dB, and 55dB, respectively. Similarly, Figs .4(d), (e), and (f) depict the variation of $\varepsilon_\sigma$ in relation to $T_n$ at the same SNR levels. The overall trends of $\sigma_T$ and $\varepsilon_\sigma$ with respect to SNR and $T_n$ are as follows: both $\sigma_T$ and $\varepsilon_\sigma$ decrease as the SNR increases, and $\varepsilon_\sigma$ exhibits a decreasing trend as $T_n$ increases. Specifically, when $T_n \geq 300$nm, $\varepsilon_\sigma$ remains consistently below 1% and is almost unaffected by NA. However, as $T_n$ decreases, $\varepsilon_\sigma$ increases sharply and becomes more significantly influenced by the NA. Under the same SNR conditions, $\varepsilon_\sigma$ for $T_n = 100$nm is nearly an order of magnitude higher compared to other $T_n$ values.

### 3.2 Source spectrum

The most distinctive feature of CSI compared to laser interferometry is its use of a broad-band light source, which produces a low-coherence envelope.

Generally, the spectral distribution of the light source could be measured, and the camera's spectral response should be considered. Assuming the spectral distribution of the light source and spectral responses of the camera are known and are shown in Fig.5. Figure 6 shows CSI signals and their Fourier magnitudes for both the spectral distribution of the light source itself, and the actual spectral distribution with also camera's spectral response into consideration.

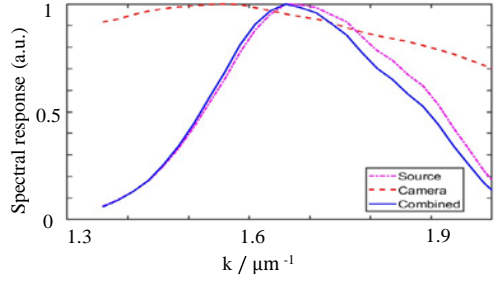

Fig. 5 Spectral distribution of light source, camera, and combined effects.

Figures 7(a) and (b) illustrate the variation of $\delta_T$ and $\varepsilon_T$ with respect to $T_n$, under the condition of deviation in the source spectrum. The overall trends are as follows: $\delta_T$ is consistently negative; the absolute value of $\delta_T$ and $\varepsilon_T$ decrease as $T_n$ increases, while they generally increase with higher NA. Specifically, for NA = 0.3 and 0.4, $\varepsilon_T$ remains approximately within the range of 0.1% to 0.2%. When NA = 0.55, $\varepsilon_T$ stabilizes around 0.4%; however, as $T_n$ decreases to 100 nm, $\varepsilon_T$ sharply increases to 1.2%.

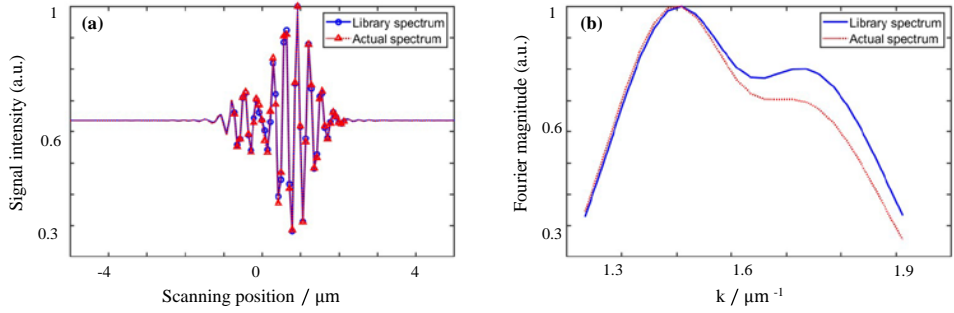

Fig. 6 CSI signals and their Fourier magnitudes using actual light spectrum and library light spectrum for 500 nm SiO$_2$/Si film structure at NA=0.55 (a) signals (b) Fourier magnitudes.

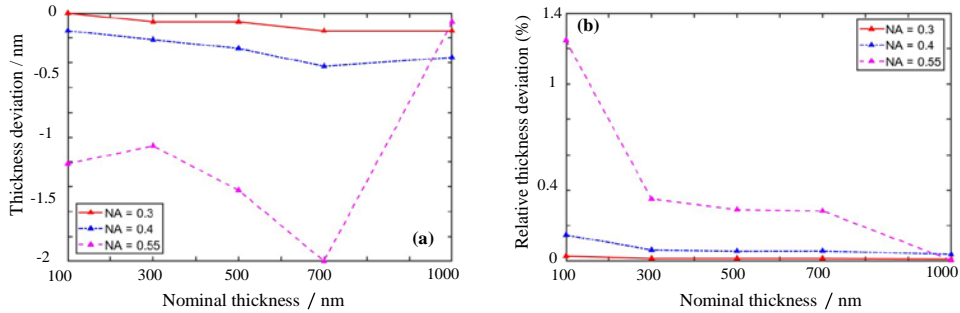

Fig. 7 Absolute and relative thickness deviations along with different film thicknesses at different NA values. (a) absolute thickness deviation (b) relative thickness deviation.

### 3.3 Numerical aperture

The actual NA value of an interferometric microscope objective can deviate from its nominal value due to design, manufacturing and assembly errors. According to existing literature on measuring NA for microscope objectives [36], the relative error ranges from −20% to 17% across various commercial products In our simulations, NA deviation is assumed within the range of −10% to 10%.

Figure 8 shows the CSI signals and their Fourier magnitudes for a 500 nm SiO$_2$/Si structure, at NA values of 0.55 and 0.539 (representing a −2% NA deviation). The comparison highlights that even a small deviation in NA can lead to notable differences in the Fourier magnitudes. This underscores the necessity of careful calibration of NA value when taking the Fourier magnitude as the signature in the model-based method [22,37].

Figures 9(a), (b) and (c) illustrate the variations of $\delta_T$ and $\varepsilon_T$ with respect to $T_n$ and NA deviations (±10%, ±5%) for NA values of 0.3, 0.4, and 0.55, respectively. The general trends can be summarized as follows: A positive NA deviation results in a positive $\delta_T$ while a negative NA deviation leads to a negative $\delta_T$. For NA = 0.3 and NA = 0.4, $\delta_T$ increases as $T_n$ increases, while $\varepsilon_T$ decreases as $T_n$ increases. Particularly, under the same conditions, $\delta_T$ and $\varepsilon_T$ for NA = 0.4 are approximately twice those for NA = 0.3, remaining around 0.5%; for NA = 0.55, when $T_n \lesssim 700$ nm, $\varepsilon_T$ fluctuates around 0.2% to 0.5%, but when $T_n$ reaches 1000 nm, $\varepsilon_T$ increases sharply with 10% NA deviation.

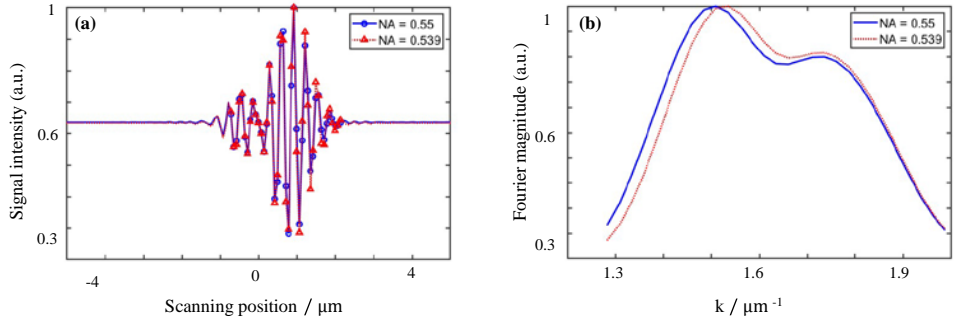

Fig. 8 CSI signals and their Fourier magnitudes for a 500 nm SiO$_2$/Si film structure at a nominal NA=0.55 and at NA=0.539 (relative error −2%) (a) signals (b) Fourier magnitudes

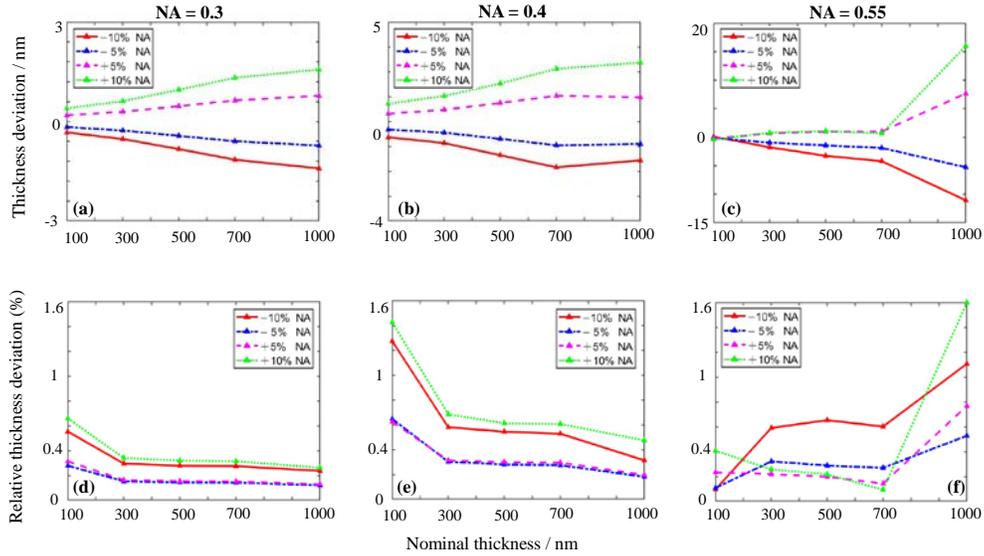

Fig. 9 Absolute and relative thickness deviations across different NAs (0.3, 0.4, 0.5) and NA deviation levels (−10%, −5%, 5%, 10%) along with different film thicknesses. (a)(d) NA = 0.3, (b)(c) NA = 0.4, and (c)(f) NA = 0.55.

## 3.4 Pupil apodization

The pupil apodization is referred to as the intensity distribution at the back-focal plane of the objective, and is typically assumed to be uniform [32-34]. However, in practical applications, the pupil apodization is often non-uniform even if Köhler illumination is used. In our simulations, the pupil apodization is assumed to be rotationally symmetrical but decreases linearly from the center to the edge of the pupil.

In Fig. 10, CSI signals and their Fourier magnitudes are illustrated for a 500nm SiO$_2$/Si structure at NA=0.55, comparing the effects of uniform apodization with non-uniform apodization, where the intensity decreases linearly from 1 at the center to 0.85 at the edge.

Figures 11(a) and (b) show the variation of $\delta_T$ and $\varepsilon_T$ with $T_n$ when there is a non-uniform intensity distribution on the pupil plane. Overall, the trends are as follows: $\delta_T$ is always negative; both the absolute value of $\delta_T$ and $\varepsilon_T$ decrease as $T_n$ increases, while they generally increase with an increase in NA. Specifically, for NA = 0.3 and NA = 0.4, $\varepsilon_T$ is mostly around 0.05% to 0.15%; when NA = 0.55, $\varepsilon_T$ remains around 0.2%, but as $T_n$ decreases to 100 nm, $\varepsilon_T$ increases quickly to 0.7%.

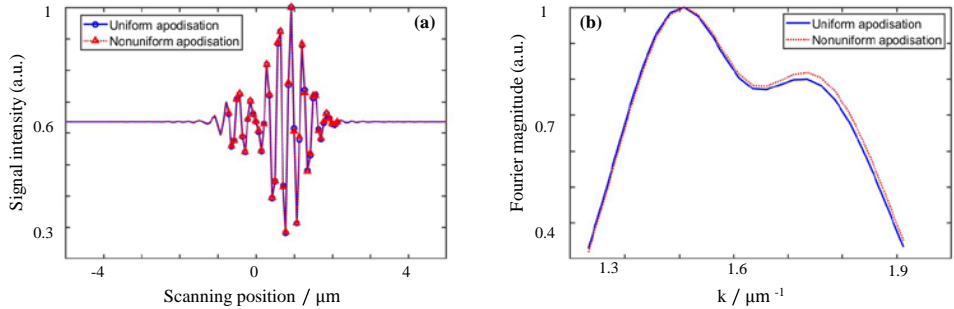

Fig. 10 CSI signals and their Fourier magnitudes for 500 nm SiO$_2$/Si film structure at NA=0.55 with uniform apodization and nonuniform apodization (a) signals (b) Fourier magnitudes.

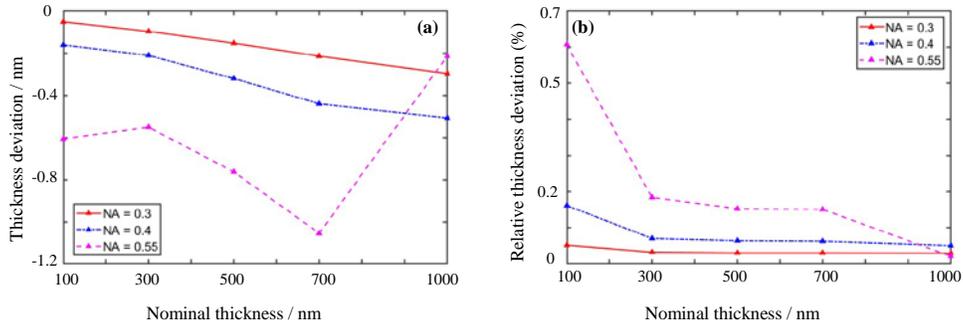

Fig. 11 Absolute and relative thickness deviations across different NA values (0.3, 0.4, 0.55) with nonuniform apodization and varying film thicknesses (a) absolute thickness deviation (b) relative thickness deviation.

## 3.5 Film refractive index

Refractive index spectra for both the film and substrate can usually be accessible from published databases. However, the actual refractive index spectrum of the film material may differ depending on the manufacturing methods and deposition conditions of the film. Typically, the refractive index spectrum may deviate by approximately 0.1–2% based on the specific manufacturing process [38]. In Fig. 12, CSI signals and their Fourier magnitudes are shown for film refractive index spectrum obtained from a public database and for a film refractive index spectrum with a 1% deviation. The conditions are also utilized in Fig. 13 which presents $\delta_T$ and $\varepsilon_T$ across different NAs.

Figures 13(a) and (b) show the variation of $\delta_T$ and $\varepsilon_T$ with $T_n$ when a 1% deviation is present in the film's refractive index spectrum.

The patterns observed are as follows: in Fig. 13(a), the $\delta_T$-thickness curve shows an approximately linear relationship, and $\delta_T$ introduced by refractive index errors is not sensitive to different NA. According to Fig. 13(b), $\varepsilon_T$ shows a slight decreasing trend as $T_n$ increases, but it remains mostly around 1% to 1.5%. This range is consistent with the magnitude of the given film refractive index error.

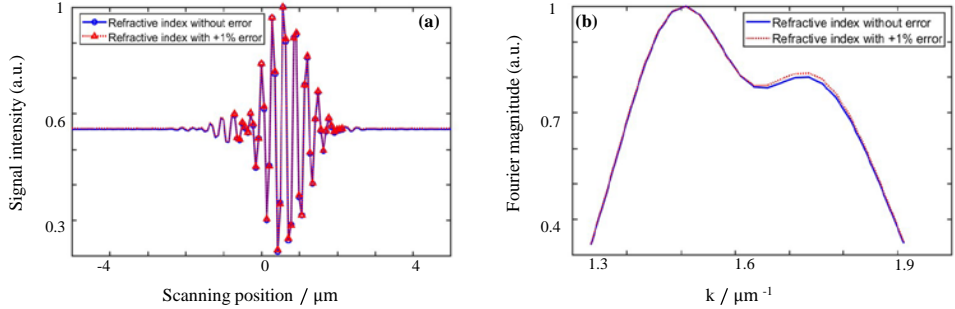

Fig. 12 CSI signals and their Fourier magnitudes for a 500 nm SiO₂/Si film structure at NA=0.55, comparing film refractive index with +1 % difference (a) signals (b) Fourier magnitudes

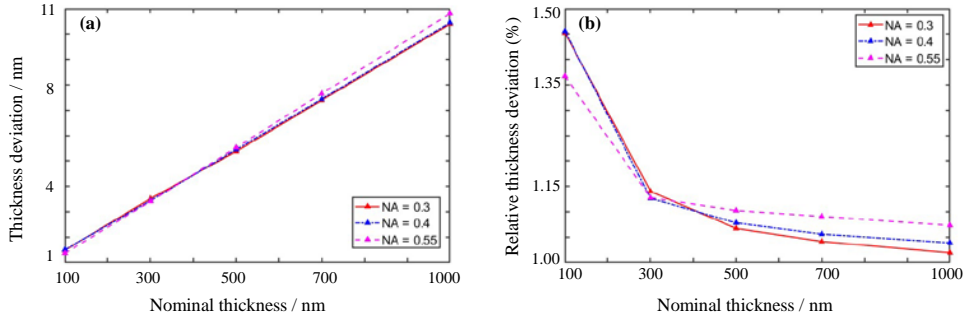

Fig. 13 Absolute and relative thickness deviation across different NAs (0.3, 0.4, 0.55) against different film thicknesses with film refractive index error. (a) absolute thickness deviation (b) relative thickness deviation

## 4. Experimental verification

### 4.1 Experimental conditions

A series of SiO2/Si film structures were analyzed using a CSI system. Interferometer microscope objectives (CF IC EPI Plan DI, Nikon, Japan) with magnifications of ×20 and ×50 were employed, as a ×10 objective is not available. These objectives have nominal NA values of 0.40 and 0.55, respectively. A model-based analysis algorithm is developed to obtain the thickness for each thin film structure, with the average thickness of each film determined as the final measured thickness. The nominal thicknesses of these films are verified using a commercial spectroscopic ellipsometry (UVISEL Plus, Horiba, Japan).

## 4.2 Result

Table 1 lists five statistical metrics $\bar{T}_m$, $\delta_T$, $\sigma_T$, $\varepsilon_T$, and $\varepsilon_\sigma$ derived from the library with nominal parameters, based on 10 repeated measurements for each film structure. $\varepsilon_\sigma$ exhibits stability when $T_n \geq 300$ nm, but increase by nearly an order of magnitude at around 100 nm, which is consistent with the simulation results. Figures 14 and 15 respectively present the linear response between $\bar{T}_m$ and $T_n$ and the $\delta_T$ values from 10 repeated measurements, for NA=0.4 and NA=0.55.

**Table 1 Statistical metrics of repeated measurements**

| $T_n$ (nm) | NA=0.4 | | | | | NA=0.55 | | | | |
|---|---|---|---|---|---|---|---|---|---|---|
| | $\bar{T}_m$(nm) | $\delta_T$(nm) | $\sigma_T$(nm) | $\varepsilon_T$(%) | $\varepsilon_\sigma$(%) | $\bar{T}_m$(nm) | $\delta_T$(nm) | $\sigma_T$(nm) | $\varepsilon_T$(%) | $\varepsilon_\sigma$(%) |
| 118.90 | 118.57 | −0.33 | 0.67 | 0.27 | 0.56 | 116.95 | −1.95 | 2.19 | 1.64 | 1.83 |
| 309.14 | 309.62 | 0.48 | 0.25 | 0.15 | 0.08 | 308.18 | −0.96 | 0.33 | 0.31 | 0.10 |
| 495.63 | 495.98 | 0.35 | 0.20 | 0.07 | 0.04 | 495.00 | −0.63 | 0.51 | 0.13 | 0.10 |
| 707.04 | 712.11 | 5.07 | 0.44 | 0.72 | 0.06 | 711.41 | 4.37 | 0.24 | 0.62 | 0.03 |
| 995.53 | 995.74 | 0.21 | 0.53 | 0.02 | 0.05 | 990.15 | −5.38 | 0.48 | 0.54 | 0.05 |

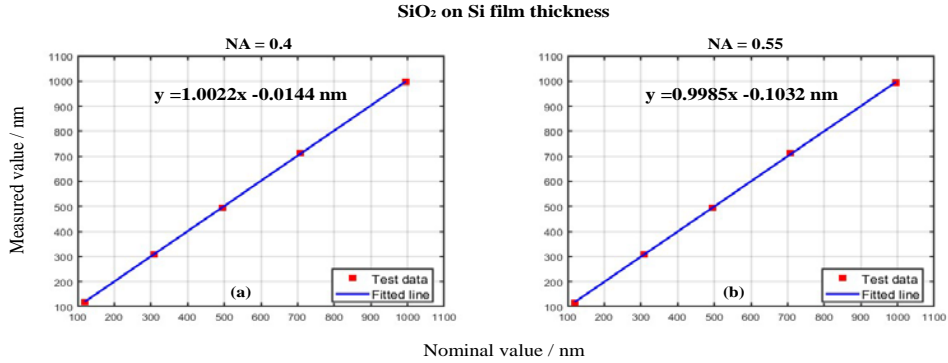

Fig. 14 Measured film thickness with CSI against SiO2/Si film structure of different film thicknesses.

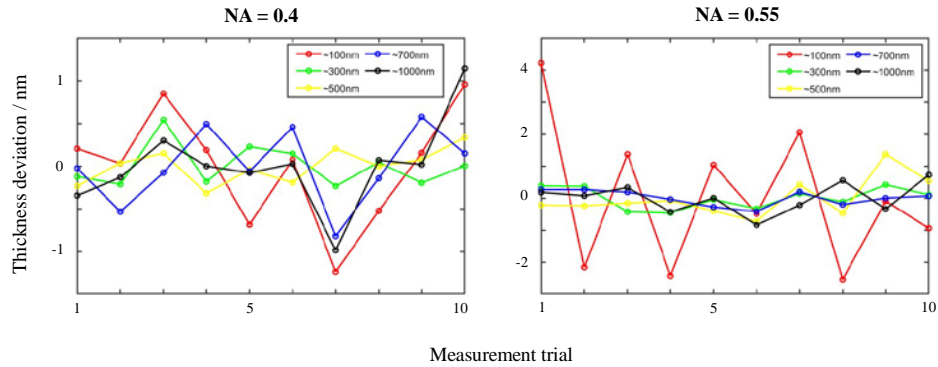

Fig. 15 Absolute thickness deviation with CSI against SiO2/Si film structure of different film thicknesses for NA =0.4 and 0.55 respectively.

In the experiments, it is not feasible to obtain $\delta_T$ by varying only one parameter while keeping all other parameters fixed at their actual values, as is typically done in simulations. Therefore, for the sensitivity analysis of the thickness deviation with respect to different instrumental parameter errors, the nominal thickness obtained from spectroscopic ellipsometry cannot serve as a reference. Instead, we use the measured thickness derived from a model-based method under nominal conditions as the reference. By applying the OAT principle, we then calculate the deviation of the measured thickness from this reference to assess the sensitivity to variations in specific parameters. This approach is reasonable, as the measured thicknesses under nominal instrumental parameters exhibit only small deviations from the nominal thicknesses, as demonstrated in Table 1.

**Table 2 Error settings in instrumental and material parameters**

| Parameters | Parameter deviation |
|---|---|
| NA | +5% |
| Apodization | 1 ~ 0.85 |
| Source spectrum | camera response considered |
| Film refractive index | +1% |

Table 2 presents the error settings in instrumental and material parameters in the experiment. The error sources are assumed as following: a +5% deviation in NA from the nominal value, non-uniform pupil apodization (1~0.85), spectral distribution of the light source including additional response measured of the camera, and a +1% deviation in the film refractive index spectrum from database values.

Figure 16 illustrates both the simulation and experiment results of the relative thickness deviations at NA= 0.4 and NA=0.55 for different parameter error case. The simulation results align well with experimental findings, showing relative thickness deviations due to errors NA errors, apodization, and source spectrum remaining below 1%. However, a 1% deviation in the film's refractive index spectrum induces at least a 1% relative deviation in the measured film thickness. Inconsistencies are also observed for thin films as thin as 100 nm or at higher NAs, such as 0.55. These inconsistencies can be attributed to the influence of additional deviations in instrumental and material parameters during experiments, as this sensitivity analysis in this work is limited to OAT local analysis.

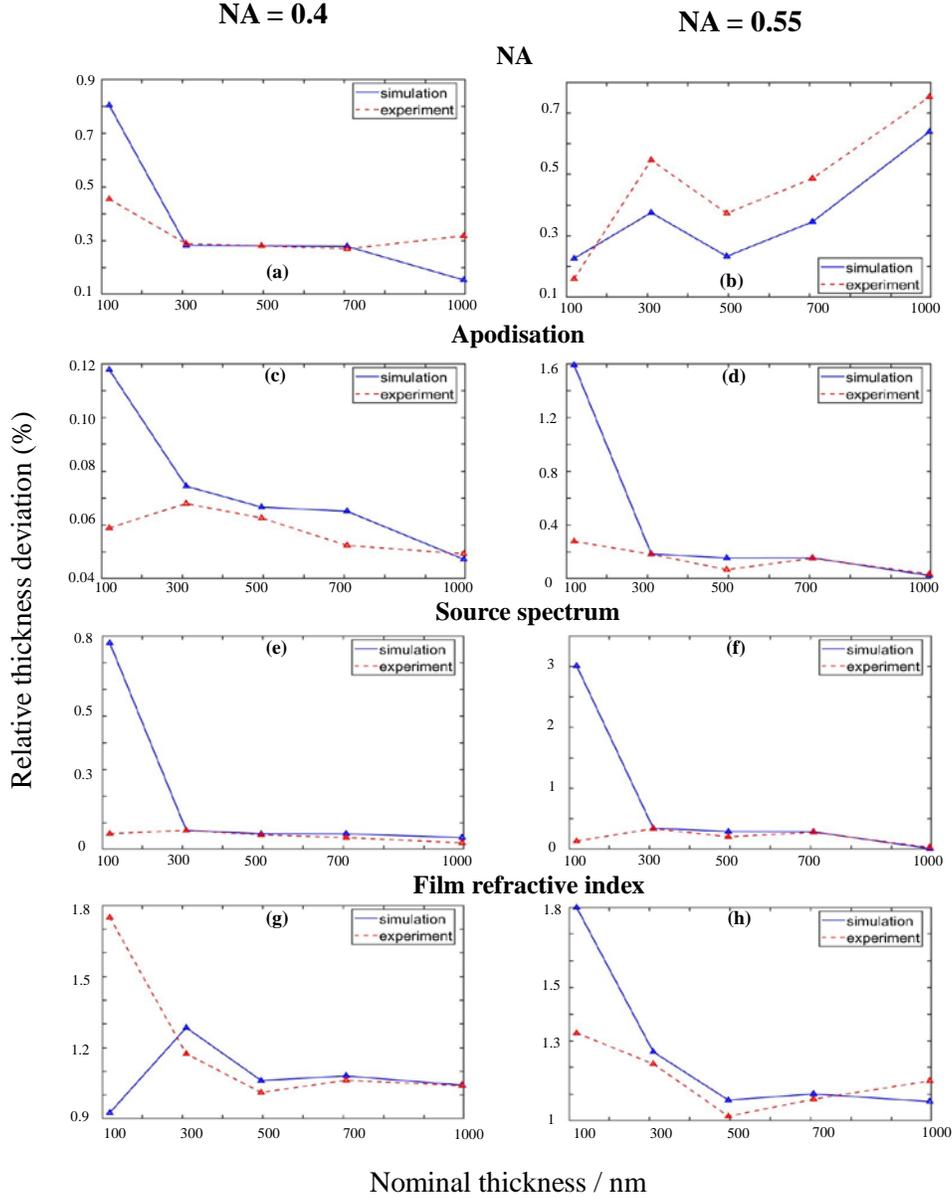

**Fig. 16** Comparisons of relative thickness deviations from different error sources in experiments and simulations against different film thicknesses for NA=0.4 and 0.55, respectively. (a)(b) for NA error, (c)(d) for apodization error, (e)(f) for source spectrum error and (g)(h) for film refractive index error.

## 5. Summary

This paper provides a quantitative analysis of the impact of inaccuracies in CSI instrumental and material parameters on film thickness measurements. Through an OAT analysis, each parameter is evaluated independently, isolating its individual effects while keeping all other parameters constant. The results show that camera noise has a negligible impact on thickness measurement accuracy for films thicker than 100 nm but becomes a critical factor for ultra-thin films below 100 nm. Both simulations and experiments reveal that errors in the NA, the apodization, and the light source spectrum have minor effects at low NAs or for films thicker than 100 nm. However, deviations in the material's refractive index have the most significant influence on the accuracy of thickness measurement, particularly for films as thin as 100 nm or CSI systems at higher NAs. This work provides guidance how to improve the accuracy of film thickness measurements using CSI.

**Funding.**

National Key R&D Program of China (2022YFE0204800); National Natural Science Foundation of China (52335010, 62105204); Strategic Priority Research Program of the Chinese Academy of Sciences, Grant No. XDA0380000

**Disclosures.**

The authors declare no conflicts of interest.

**Data availability.**

Data underlying the results presented in this paper are not publicly available at this time but may be obtained from the authors upon reasonable request.